\documentclass[10pt]{iopart}
\usepackage[utf8]{inputenc}
\usepackage{iopams}
\usepackage{graphicx}
\usepackage{cite}
\newcommand{\text}[1]{\mathrm{#1}}

\begin{document}

\title{Wakefield decay in a radially bounded plasma due to formation of electron halo}

\author{R.I. Spitsyn}
%\address{Budker Institute of Nuclear Physics, Novosibirsk, 630090, Russia}
\address{Novosibirsk State University, Novosibirsk, 630090, Russia}
\ead{R.I.Spitsyn@inp.nsk.su}
\author{K.V. Lotov}
%\address{Budker Institute of Nuclear Physics, Novosibirsk, 630090, Russia}
\address{Novosibirsk State University, Novosibirsk, 630090, Russia}
\ead{K.V.Lotov@inp.nsk.su}

\vspace{10pt}
\begin{indented}
\item[]\today
\end{indented}

\begin{abstract}
There is a new effect that can limit the lifetime of a weakly nonlinear wakefield in a radially bounded plasma. If the drive beam is narrow, some of the plasma electrons fall out of the collective motion and leave the plasma radially, forming a negatively charged halo around it. These electrons repeatedly return to the plasma under the action of the charge separation field, interact with the plasma wave and cause its damping. The lowest-energy halo electrons take the energy from the wave more efficiently, because their trajectories are bent by the plasma wave towards the regions of the strongest acceleration. For correct accounting of the wave damping in simulations, it is necessary and sufficient to take the simulation window twice as wide as the plasma.
\end{abstract}

\vspace{2pc}
\ioptwocol

\section{Introduction}\label{s1}

Plasma wakefield acceleration is a novel technique that promises to increase the electron energy available to 
research laboratories \cite{RMP81-1229,RMP90-035002,RAST9-63,RAST9-85}. The wave that accelerates particles must be excited in the plasma by a driver: a laser, an electron or a proton beam. At present, only proton beams have sufficient energy to boost electrons to TeV energies in a single accelerating stage \cite{NatPhys5-363,PoP18-103101}. 

Proton beams in modern high-energy synchrotrons are much longer than the plasma wavelength and can efficiently drive the wave only after micro-bunching, that is, splitting into a train of short bunches either before \cite{RuPAC16-303} or inside the plasma \cite{PRL104-255003,PoP22-103110,PoP22-123107}. The micro-bunches then resonantly excite the plasma oscillations over many periods. During growth, the wave remains linear or weakly nonlinear in the sense that the relative change of the wakefield period is small. Therefore, it is important to know how long the wakefield can remain regular and what physical effects limit its lifetime.

There are several effects that can destroy a plasma wave of moderate amplitude: nonlinear elongation of the wave period \cite{JETP3-696,PRL29-701,PF30-904,PRE60-6210,PoP20-083119}, ion motion \cite{PRL86-3332,PoP10-1124,PRL109-145005,PoP21-056705,PoP25-103103}, and transverse plasma non-uniformity \cite{PoP4-1145,NIMA-410-469,PoP6-591,PoP23-013109}. In this paper, we describe one more effect that occurs in a radially bounded plasma.
If the drive beam is narrow, the wavefronts of the excited wakefield quickly become curved, and trajectories of some plasma electrons intersect. These electrons fall out of the collective motion and leave the plasma radially, which by itself does not lead to a catastrophic wave damping. However, these electrons form a negatively charged halo around the plasma and repeatedly return to the wakefield region under the action of the charge separation field \cite{PRL112-194801}. The returning electrons interact with the plasma wave and cause its damping, if there are many of them.

The effect was noticed while simulating the AWAKE experiment at CERN \cite{NIMA-829-3,NIMA-829-76,PPCF60-014046}. In AWAKE, a 400\,GeV proton beam self-modulates in the plasma \cite{PRL122-054801,PRL122-054802,arXiv:2007.14894} and excites the wakefield \cite{PRAB23-081302}, which then accelerates the externally injected witness electrons \cite{Nat.561-363,PTRSA377-20180418}. Agreement between simulations and measurements is only achieved for wide simulation windows that correctly accounts for the halo electrons \cite{PPCF-125023,arXiv:2010.05715}. In a narrow window, the backflow of halo electrons is suppressed, the wakefield persists much longer than in a wide window, and simulation results disagree with experimental observations.

\begin{table}[tb]
\caption{Parameters for numerical simulations.}\label{tab:t1}
\begin{indented}
\item[]\begin{tabular}{@{}ll}\br
Parameter, notation & Value \\ \mr
Plasma density, $n_0$ & $1.965 \times 10^{14}\,\text{cm}^{-3}$ \\
Plasma skin depth, $c/\omega_p$, & 380\,$\mu$m \\
Plasma radius, $R_p$ & 1.5\,mm ($\approx 4 c/\omega_p$) \\
Plasma ion-to-electron mass ratio, $M_i$ & 157\,000\\
Plasma initial temperature, $T_e$ & 0\,eV \\
Beam population, $N_b$ & $2.86\times 10^{11}$ \\
Beam energy, $W_b$ & 400\,GeV \\
Beam energy spread, $\delta W_b$ & 0.35\% \\
Beam length, $\sigma_{z}$ & 6.57\,cm ($\approx 170 c/\omega_p$) \\
Beam radius, $\sigma_{r}$ & 200\,$\mu$m ($\approx 0.5 c/\omega_p$) \\
Normalized beam emittance, $\epsilon_b$ & 3.6\,mm\,mrad \\
Seed pulse position, $\xi_s$ & 3.75\,cm ($\approx 100 c/\omega_p$) \\
\br
\end{tabular}
\end{indented}
\end{table}

We study the details of plasma wave suppression by halo electrons numerically using the quasistatic axisymmetric code LCODE \cite{PRSTAB6-061301, NIMA829-350} and focus on the same set of parameters that was analyzed in reference~\cite{PPCF-125023} as the baseline case (table~\ref{tab:t1}). The self-modulation is seeded by a laser pulse co-propagating with the proton beam at the distance $\xi_s \approx 0.6\sigma_{z}$ ahead of its center. The simulation window encloses the entire proton beam and moves with the speed of light $c$ in the positive $z$ direction. It has a size of $r_w = 100 c/\omega_p$ in the radius and up to $720 c/\omega_p$ in the longitudinal coordinate $\xi = z - ct$, where $\omega_p = \sqrt{4 \pi n_0 e^2 / m}$ is the plasma frequency, $n_0$ is the plasma density, $e$ is the elementary charge and $m$ is the electron mass. The coordinate $\xi=0$ corresponds to the seed laser pulse position. The grid size is $0.01 c/\omega_p$ in the $r$ and $\xi$ directions, and there are 10 or more (for drawing electron trajectories) radially weighted plasma particles of each sort per cell. We simulate the beam propagation to complete micro-bunching (at $z=4$\,m) with steps $\Delta z = 200 c/\omega_p \approx 7.6$\,cm and then study the wakefield excited by this beam. We characterize the wakefield strength by the wakefield potential $\Phi$, which shows both radial and longitudinal forces acting on an axially moving unit charge:
\begin{equation}\label{Phi}
E_r - B_{\phi} = -\frac{\partial \Phi}{\partial r}, \qquad E_z = -\frac{\partial \Phi}{\partial \xi}.
\end{equation}

In section~\ref{s2}, we describe how the electron halo appears and what are its properties. In section~\ref{s3}, we analyze the energy exchange between the halo and the plasma wave and show that this is the main cause of wave damping. Then, in section~\ref{s5}, we consider the interaction of individual halo electrons with the wave and show that the energy preferably flows from the wave to halo because of bending of electron trajectories by the wave. Finally, in section~\ref{s6}, we summarize the main findings.

\begin{figure}[tb]
	\centering
	\includegraphics{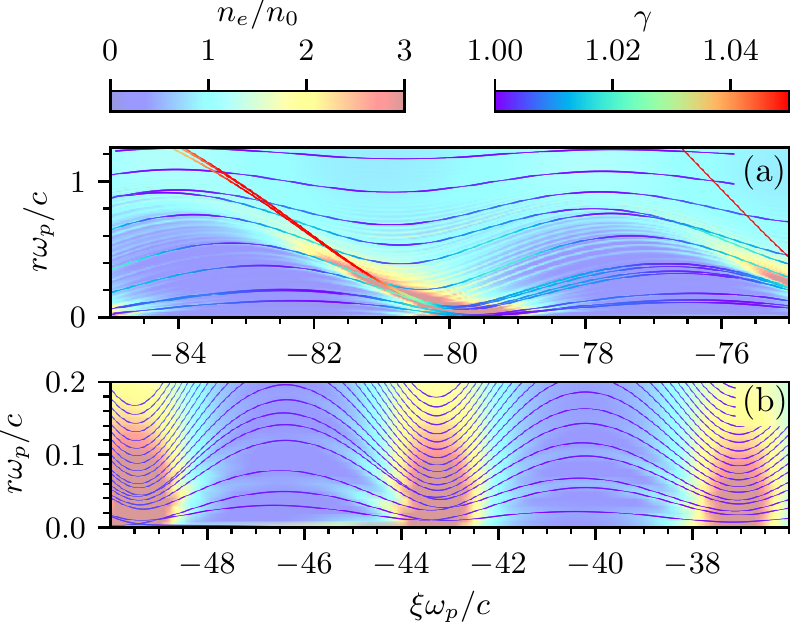} 
	\caption{Typical electron trajectories plotted against the background of the electron density $n_e$: (a) wavebreaking events followed by the appearance of halo electrons, (b) intersections of trajectories that do not lead to wavebreaking. Color of the lines show the relativistic factor $\gamma$ of electrons.}\label{img:Intersections}
\end{figure}
\begin{figure}[tb]
	\centering
	\includegraphics{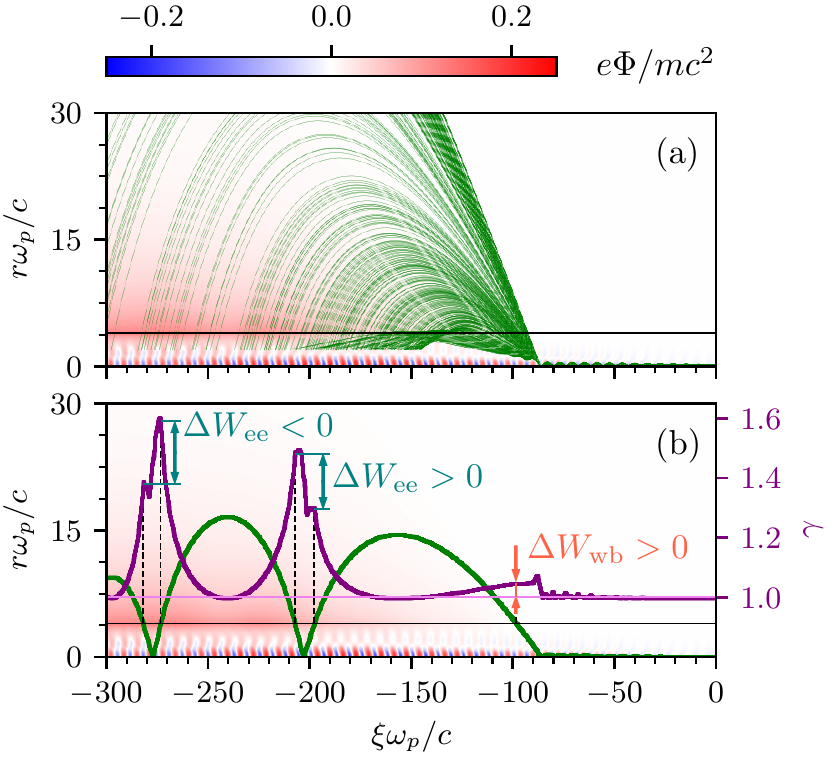} 
	\caption{Selected trajectories of halo electrons (green lines) plotted against the background of the wakefield potential $\Phi$ (color maps). (a) A family of halo electrons appearing during one wakefield period (in the interval $-86.5 < \xi \omega_p/c < -85.5$). 
	The trajectories are drawn up to the first approach to the axis so that they do not cover the potential map near the axis. (b) A single plasma electron and its relativistic factor $\gamma (\xi)$ (purple line). Thin  horizontal black lines are the plasma boundary.}\label{img:Trajectories}
\end{figure}

\section{Motion of halo electrons}\label{s2}

Plasma electrons drop out of the wave if the trajectories of inner and outer electrons intersect at a high ridge of electron density %\cite{newGorn} 
(figure~\ref{img:Intersections}(a)). The former inner electron continues its radial motion and eventually becomes a halo electron. Other intersections of trajectories, which are also possible in the wave, do not lead to the emission of electrons (figure~\ref{img:Intersections}(b)). The halo electrons appear in certain phases of the wakefield, and then their trajectories diverge, like a jet from a watering hose, and return to the plasma in a wide interval of $\xi$ (figure~\ref{img:Trajectories}(a)). The distance traveled by an electron outside the plasma depends on the energy $\Delta W_\text{wb}$ acquired during the wavebreaking and further passing through the plasma column (figure~\ref{img:Trajectories}(b)). There is no phase correlation between successive points of exit from and entry into the plasma.

\begin{figure}[tb]
	\centering
	\includegraphics{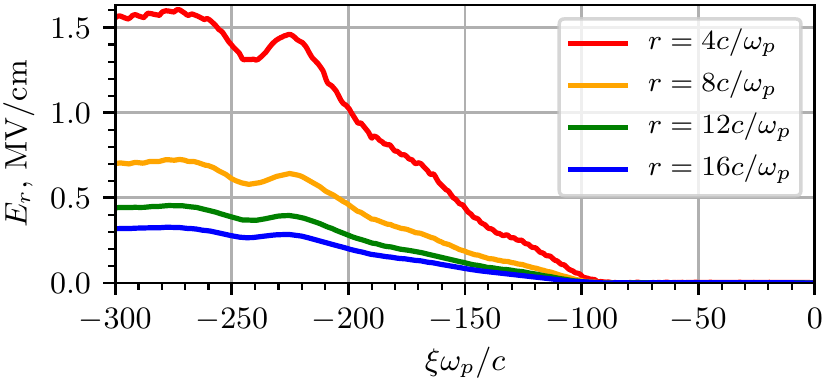} 
	\caption{Radial electric field $E_r (\xi)$ at different radii outside the plasma column.
	}\label{img:ErOutside}
\end{figure}

Halo electrons can intersect the plasma several times, each time changing the energy by $\Delta W_\text{ee}$ (figure~\ref{img:Trajectories}(b)). If the electron crosses the accelerating phase of the wave, it takes the energy from the wave ($\Delta W_\text{ee} > 0$), and returns it otherwise  ($\Delta W_\text{ee} < 0$). The radial electric field around the plasma column grows as the number of halo electrons increases (figure~\ref{img:ErOutside}). For this reason, the electrons return to the plasma with a higher energy than when they leave (figure~\ref{img:Trajectories}(b)).

\section{Energy balance}\label{s3}

\begin{figure}[tb]
	\centering
	\includegraphics{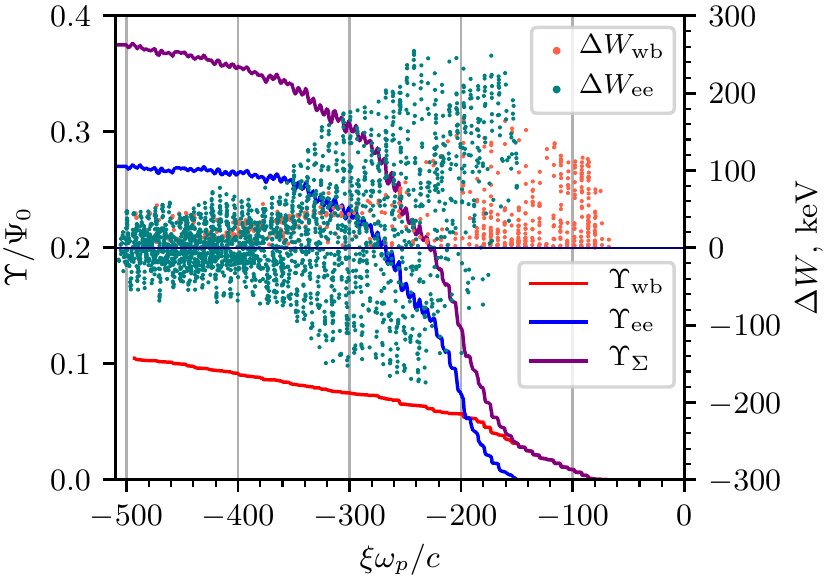} 
	\caption{The energy exchange between halo electrons and the plasma wave. The red dots show the energies $\Delta W_\text{wb}$ of the electrons ejected from the plasma for the first time; these values are always positive. The teal dots correspond to the energy changes $\Delta W_\text{ee}$ when electrons pass through the plasma column. The red, blue and purple lines are the values of $\Upsilon_\text{wb}$, $\Upsilon_\text{ee}$ and $\Upsilon_{\Sigma}$, respectively.}\label{img:EtakenR100}
\end{figure}

Damping of a plasma wave is always associated with a rapid loss of its energy. To understand the role of halo electrons in this process, we quantitatively characterize the energy exchange between them and the plasma wave. Let us correlate the amount of energy $\Delta W$ received or returned by a halo electron in a single act of energy exchange and the coordinate $\xi_e$ at which the electron crosses the axis (for $\Delta W = \Delta W_\text{ee}$) or where the electron radius has the last local minimum before leaving the plasma (for $\Delta W = \Delta W_\text{wb}$) (figure~\ref{img:EtakenR100}). These energy changes can then be sequentially summed up, starting from the beam head, to obtain the total energies directly transferred from the wave to the halo:
\begin{equation}{\label{E_separate}}
    \Upsilon_\text{wb} \left( \xi \right) = \sum_{\xi_e > \xi} \Delta W_\text{wb}(\xi_e), \ \
    \Upsilon_\text{ee} \left( \xi \right) = \sum_{\xi_e > \xi} \Delta W_\text{ee}(\xi_e),
\end{equation}
\begin{equation}{\label{E_sum}}
    \Upsilon_{\Sigma} = \Upsilon_\text{wb} + \Upsilon_\text{ee}.
\end{equation}
The summations in the formula (\ref{E_separate}) are carried out over electrons that enter the simulation window per unit of time, so the quantities $\Upsilon_\text{wb}$, $\Upsilon_\text{ee}$, and $\Upsilon_{\Sigma}$ have the dimension of the energy flux. We use 
\begin{equation}\label{eq3}
\Psi_0 = \frac{m^2 c^5}{4 \pi e^2}
\end{equation}
as the unit for them (figure~\ref{img:EtakenR100}). Note that $\Upsilon_{\Sigma}$ does not represent the actual energy content of the halo, as the halo electrons also change their energy while interacting with the electric field around the plasma (figure~\ref{img:Trajectories}(b)).

\begin{figure}[tb]
	\centering
	\includegraphics{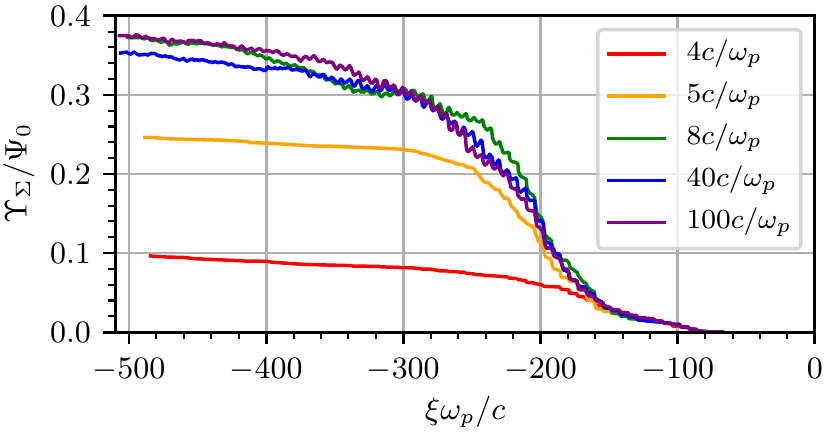} 
	\caption{The energy $\Upsilon_{\Sigma} (\xi)$ transferred from the plasma wave to halo electrons in numerical simulations with windows of different widths $r_w$ (indicated in the legend). The lowest (red) line corresponds to the case when the wall is located almost at the plasma boundary; therefore, its behavior is mainly determined by the energy $\Upsilon_\text{wb} (\xi)$ acquired by the halo electrons during the wavebreaking.}\label{img:EtakenRWalls}
\end{figure}

In numerical simulations, the behavior of $\Upsilon_{\Sigma} (\xi)$ depends on the radius of the simulation window (figure~\ref{img:EtakenRWalls}). The outer boundary of the LCODE simulation window acts as a perfectly conducting wall that reflects plasma electrons and absorbs their energy. If this wall is located close to the plasma, it suppresses halo formation and, as a consequence, the wave damping \cite{PPCF-125023}. Choosing a very distant wall, like in real experiments, is undesirable because of the large computational resources required. A compromise is possible if the simulation window is about twice wider than the plasma. In this case, the boundary of the simulation window still disturbs the trajectories of some halo electrons, but the overall effect of the halo on the wave is almost the same as in a very wide window. We discuss the reasons for this in section~\ref{s5}. Thus, the simulation window width of $100 c/\omega_p$ is chosen with a large margin.

\begin{figure}[tb]
	\centering
	\includegraphics{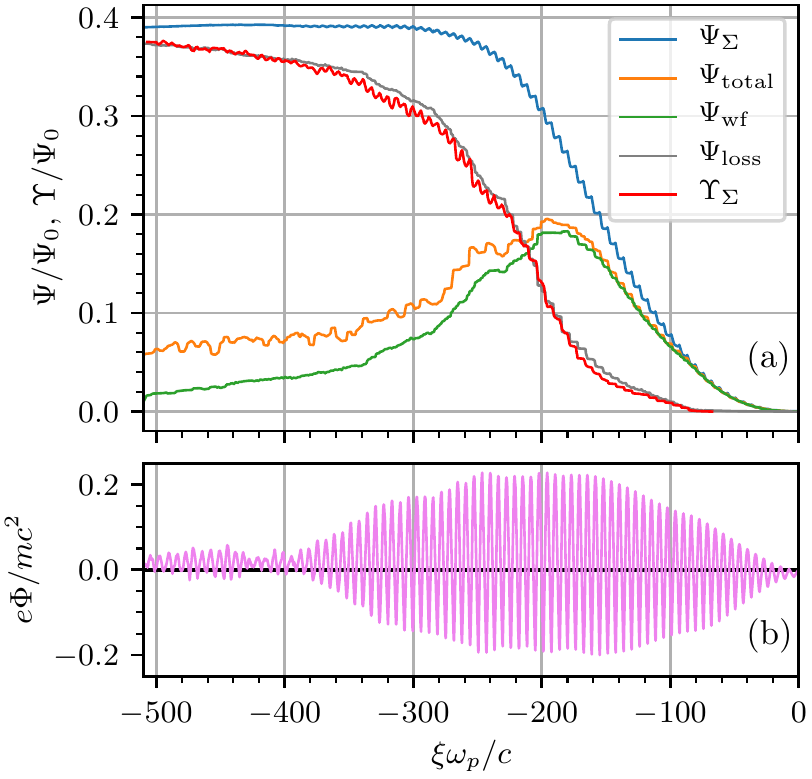} 
	\caption{(a) Energy fluxes as functions of the longitudinal coordinate $\xi$. (b) The wakefield potential $\Phi (\xi)$ on the axis to show the regions of wave growth and decay.}\label{img:Efluxes} 
\end{figure}

To understand how the plasma wave loses its energy, we introduce the energy fluxes in the co-moving window \cite{PRE69-046405, PoP25-103103}. The total energy flux $\Psi (r_m, \xi)$ is the sum of the electromagnetic part
\begin{equation}\label{Psi_em}
\Psi_\text{em} = \int_0^{r_m} \left( \frac{c}{8 \pi} \left( E^2 + B^2 \right) - \frac{c}{4 \pi} \left[ \vec{E} \times \vec{B}\right]_z \right) 2 \pi r \, dr,
\end{equation}
and the kinetic energy part
\begin{equation}\label{Psi_p}
\Psi_\text{p} = \sum_j \left( \gamma_j - 1 \right) m_j c^2,
\end{equation}
where the summation is over plasma particles (having mass $m_j$ and relativistic factor $\gamma_j$) that intersect the circle of radius $r_m$ in unit time. One can imagine that a continuous stream of plasma particles enters the simulation window, acquires some energy in some places and transfers it to others, forming energy flows. The electromagnetic field created by the plasma and beam particles also transfers energy and contributes to the energy flux. Plasma particles that reach the upper boundary of the simulation window transfer their energy to it. This energy, integrated from the beam head to some $\xi$, forms the energy flux to the ``wall'' $\Psi_\text{wall} (\xi)$. The sum 
\begin{equation}\label{Psi_sigma}
\Psi_\Sigma = \Psi_\text{wall} + \Psi (r_w, \xi), 
\end{equation}
where the total flux $\Psi$ is calculated over the entire simulation window, characterizes the total energy taken from the driver beam. It grows as the beam excites the wave and remains constant in the intervals of $\xi$ where there is no energy exchange between the beam and the plasma (figure~\ref{img:Efluxes}).

\begin{figure}[tb]
	\centering
	\includegraphics{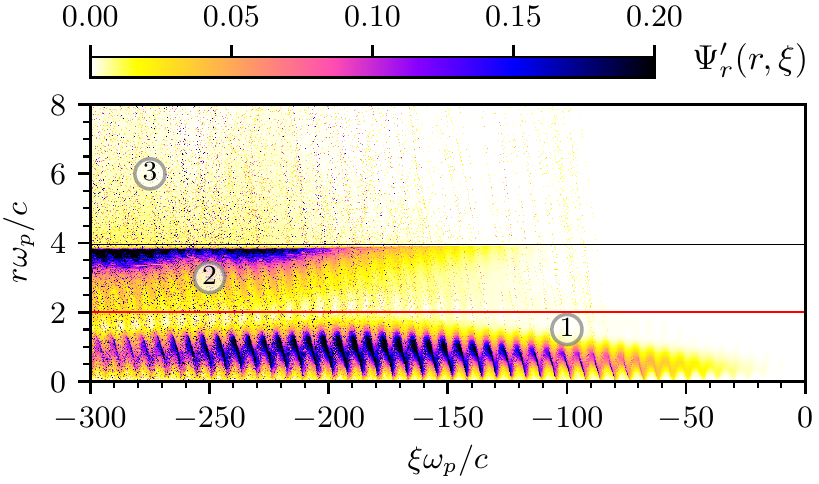} 
	\caption{Energy flow map defined by $\Psi_r' (r, \xi)$. Circled numbers mark the features discussed in the text. The black horizontal line at $r = R_p = 3.96c/\omega_p$ is the plasma boundary, and the red line at $r = r_\text{wf} = 2c/\omega_p$ separates the wave region from the rest of the plasma.}\label{img:Sf2map}
\end{figure}

The derivative
\begin{equation}\label{Sz}
\Psi_r' (r, \xi) = \frac{\partial \Psi (r, \xi)}{\partial r} 
\end{equation}
allows us to visualize areas of the simulation window that contain energy (figure~\ref{img:Sf2map}). We see the energy of the plasma wave (feature~1), the energy of the return current that appears near the plasma boundary to compensate the current of halo electrons (feature~2), and the energy of halo electrons that penetrate everywhere after the wave begins to break (feature~3). Since the group velocity of the plasma wave is very low, the wave energy remains where it was released by the beam. To calculate this energy as accurately as possible, we define the radius $r_\text{wf} = 2c/\omega_p$ of the wakefield region using the figure~\ref{img:Sf2map}. However, the total flux 
\begin{equation}\label{Psy_full}
\Psi_\text{total} = \Psi (r_\text{wf}, \xi)
\end{equation}
overestimates the wave energy, since the halo electrons that cross the wakefield region contribute to the energy flux being not a part of the plasma wave. We can improve the expression for the wave energy by noting that in a linear plasma wave, the average energy is equally distributed between the kinetic energy of particle motion and the electric field. The doubled flux of the electromagnetic energy gives us the best expression for the energy contained in the plasma wave:
\begin{equation}\label{Psy_wf}
\Psi_\text{wf} = 2 \Psi_\text{em} (r_\text{wf}, \xi).
\end{equation}
The difference between $\Psi_\text{total}$ and $\Psi_\text{wf}$ appears when the halo electrons return to the plasma wave region (figure~\ref{img:Efluxes}); before this, the lines go alike.

The difference
\begin{equation}\label{Psy_loss}
\Psi_\text{loss} = \Psi_\Sigma - \Psi_\text{wf}
\end{equation}
characterizes the energy losses from the plasma wave through all possible channels. It coincides with the energy $\Upsilon_\Sigma$ transferred to the halo electrons (figure~\ref{img:Efluxes}), proving this to be the dominating channel of wakefield energy loss and the main cause of wave damping. The halo electrons then share this energy with the electromagnetic field around the plasma and with the directionally moving plasma electrons.

\section{Lensing by the wakefield}\label{s5}

When crossing the axis, the halo electrons take the energy from the wave on average (figure~\ref{img:EtakenR100}). Amounts of energy transferred to or taken from a single electron are comparable, but the number of accelerated electrons is greater, at least when the electrons cross the axis for the first time (at $\xi \sim -200c/\omega_p$ in figure~\ref{img:EtakenR100}). Figure~\ref{img:Trajectories}(a) shows that the return points of halo electrons to the plasma are smoothly distributed over~$\xi$. Consequently, the asymmetry between accelerated and decelerated electrons arises when they interact with the plasma wave. 

\begin{figure}[tb]
	\centering
	\includegraphics{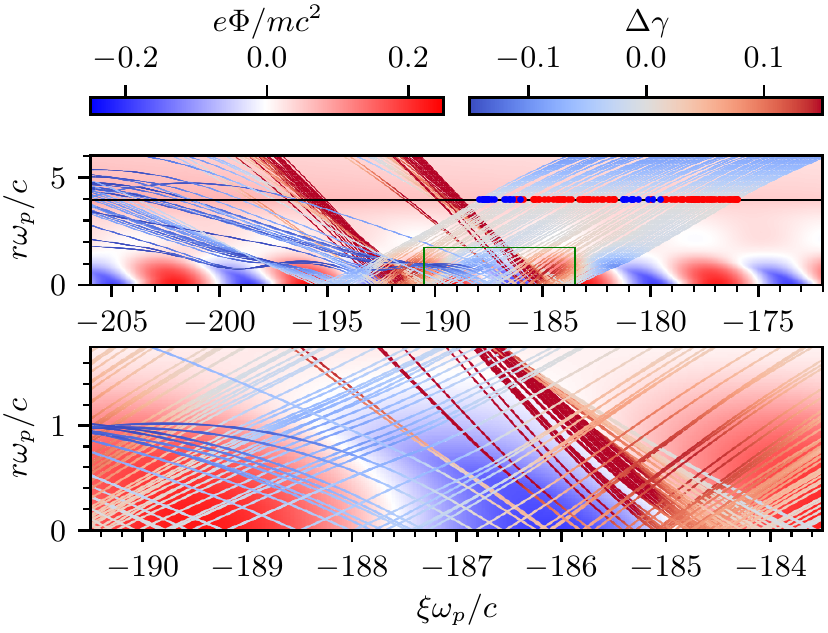} 
	\caption{Trajectories of electrons entering the plasma within two wave periods at approximately the same angles, plotted against the background of the wakefield potential $\Phi$. The bottom picture is an enlarged fragment of the top picture, marked with a green rectangle. The black horizontal line in the top picture is the plasma boundary. The red and blue dots on the trajectories at the points of their entry into the plasma indicate a positive or negative change in energy when electrons pass through the plasma. The color of the trajectories show the change of the electron relativistic factor $\Delta \gamma$ with respect to its value at the plasma boundary.}\label{img:Lensing180}
\end{figure}

To understand the reasons for the asymmetry, we consider the trajectories of halo electrons that enter the plasma at almost the same angles near the point where $\Upsilon_\text{ee}$ becomes nonzero (figure~\ref{img:Lensing180}). Initially parallel trajectories deviate from straight lines under the action of the wakefield. The force arises mainly from the electric field, since the magnetic field in a weakly nonlinear Langmuir wave is small. The magnetic field of the proton beam in the considered region is also small, because this is the field of individual micro-bunches, while the electric field is created by coherent contributions from tens of micro-bunches. Consequently, the wakefield potential is almost the same as the electric potential. Regions of negative potential (red areas in figure~\ref{img:Lensing180}) are potential "wells" for electrons, which attracts them and bend their trajectories towards regions of strongest acceleration located at the highest potential gradients. And vice versa, the blue areas of positive potential defocus the electrons away from deceleration regions. In the laboratory frame, the halo electrons move predominantly forward (in the positive $z$ direction) \cite{PRL112-194801}, so their interaction with the wave is long and efficient. We can envisage this process as if properly phased witness bunches are taking energy from the wave.

\begin{figure}[tb]
	\centering
	\includegraphics{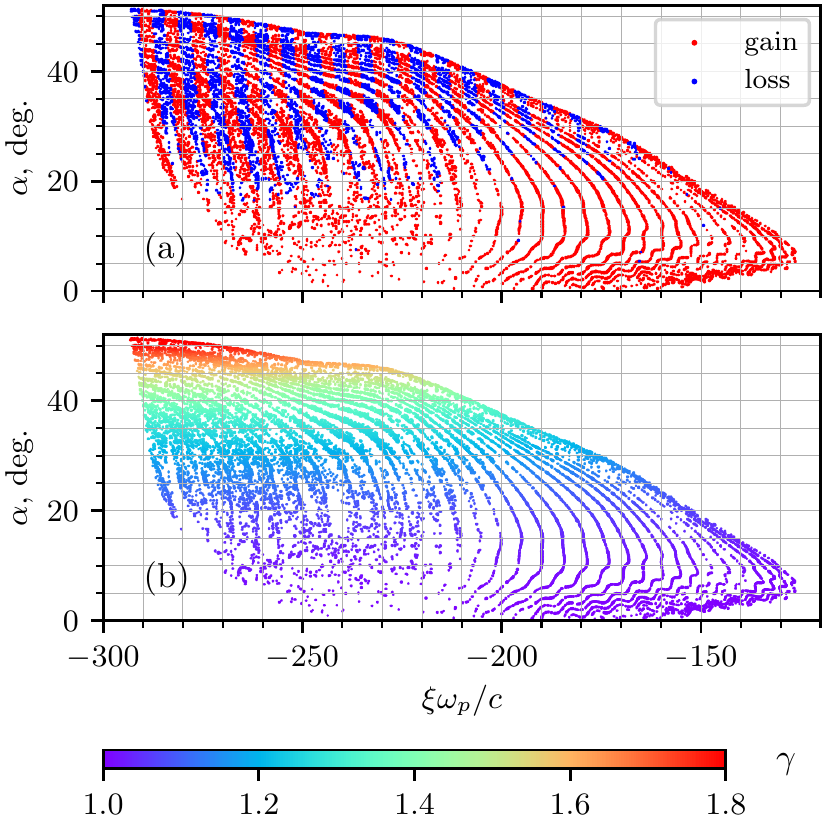} 
	\caption{Correlation between the angle $\alpha$ at which the halo electrons enter the plasma, the entry coordinate $\xi$, and (a) the energy gain (red points) or loss (blue points) when passing through the plasma, and (b) the relativistic factor of electrons $\gamma$ (color) at the entry point.}\label{img:Wings}
\end{figure}
\begin{figure}[tb]
	\centering
	\includegraphics{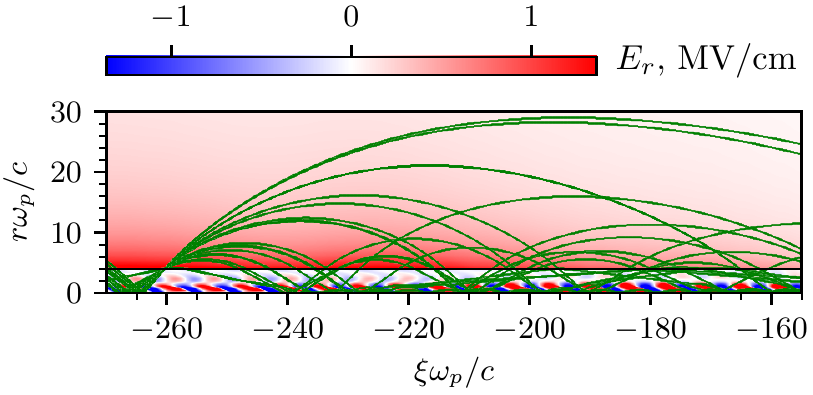} 
	\caption{A family of electron trajectories entering the plasma at $\xi \approx -260c/\omega_p$ (green lines), plotted  against the background of the radial electric field $E_r$ (color map). The black horizontal line is the plasma boundary.}\label{img:Lensing260}
\end{figure}
\begin{figure}[tb]
	\centering
	\includegraphics{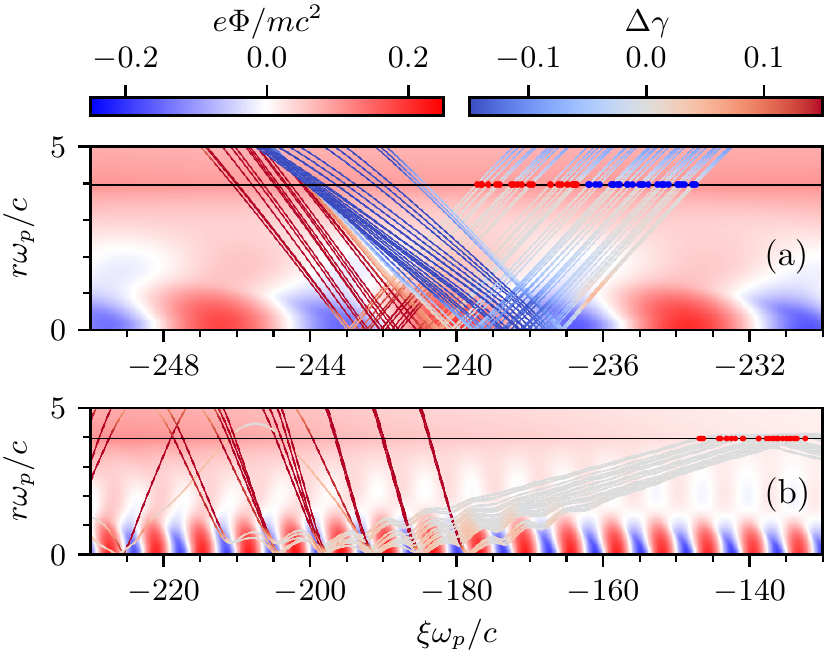}
	\caption{ Trajectories of high-energy (a) and low-energy (b) halo electrons entering the plasma at approximately the same angles, plotted against the background of the wakefield potential~$\Phi$. The black horizontal lines show the plasma boundary. The red and blue dots on the trajectories at the points of their entry into the plasma indicate a positive or negative change in energy when electrons pass through the plasma. The color of the trajectories show the change of the electron relativistic factor $\Delta \gamma$ with respect to its value at the plasma boundary. The axis ratio in fragment (b) is 1:4 to make the trajectories better visible.}\label{img:LensingAngles}
\end{figure}

The question now arises why the asymmetry of energy exchange is the strongest when the electrons just begin to return to the plasma (figure~\ref{img:EtakenR100}). To answer, let us note the correlation between the asymmetry and the angle at which the electron enters the plasma in the co-moving frame (figure~\ref{img:Wings}(a)). The smaller the angle, the stronger the asymmetry. Almost all electrons entering the plasma at angles less than $20^\circ$ gain energy. The entry angle, in turn, correlates with the electron energy (figure~\ref{img:Wings}(b)): higher-energy electrons enter at larger angles. This is because the entry angle is determined by the size of the arc along which the electron moves outside the plasma (figure~\ref{img:Lensing260}). A larger arc obviously means that the electron left the plasma with a larger radial momentum, which became even larger on return because of the electric field around the plasma, which increases with increasing $|\xi|$ (figure~\ref{img:ErOutside}). High-energy electrons propagate in the plasma along almost straight lines without being significantly deflected by the wakefield (figure~\ref{img:LensingAngles}(a)), with the number of accelerated and decelerated electrons being approximately the same. The contributions of these electrons cancel out when averaged over the plasma period, and we observe the constancy of $\Upsilon_{ee}$ at $\xi < -350c / \omega_p$ in figure~\ref{img:EtakenR100}. The trajectories of low-energy electrons are strongly bent by the wakefield and cross the axis exclusively in the accelerating phase of the wave (figure~\ref{img:LensingAngles}(b)). These electrons are the earliest to return to the plasma after the wavebreaking (figure~\ref{img:Trajectories}(a)), and their contribution to wave damping is the strongest. Since the low-energy electrons do not travel too far from the plasma, correct accounting for their contribution does not require a very wide simulation window (figure~\ref{img:EtakenRWalls}).

It is interesting to note that electrons change both longitudinal and transverse momentum when crossing the plasma wave. Let us consider a ``red'' trajectory of an accelerated electron from figure~\ref{img:LensingAngles}(a) as an example. The electron first falls into the potential well from the side and increases its radial momentum. Then the electron climbs the potential hump, gaining longitudinal momentum. Finally, it goes down from the potential hump and gains additional radial momentum. For decelerated electrons, all changes in momentum have the opposite sign.

\section{Summary}\label{s6}

To sum up, halo electrons that appear around a radially bounded plasma as a result of wavebreaking can cause a quick decay of the wakefield. Passing through the plasma, these electrons act like witness beams, taking away the energy of the plasma wave. The most efficient energy exchange occurs with the lowest-energy halo electrons, because their trajectories are bent by the plasma wave towards the regions of strongest acceleration. The low-energy halo electrons do not travel too far from the plasma boundary. Therefore, for correct accounting of the wave damping in simulations, it is sufficient to take the simulation window twice wider than the plasma. In this paper, we described the mechanisms of halo formation and wave decay due to the acceleration of halo electrons and left the study of the parameters at which this mechanism dominates for future works.

\ack
This work was supported by the Russian Science Foundation, project 20-12-00062. Simulations were performed on HPC-cluster ``Akademik V.M.\,Matrosov'' \cite{matrosov}.

\end{document}